\documentclass[english,aps,preprint]{revtex4}

\usepackage{geometry}
\usepackage{babel}
\usepackage{graphics}
\usepackage{subfigure}
\usepackage{times}
\usepackage{latexsym}

\begin{document}

\title{New geometries for high spatial resolution Hall probes}

\author{H. Guillou}
\email{herve.guillou@ujf-grenoble.fr}
\affiliation{CRTBT-CNRS, laboratoire associé à l'université Joseph Fourier, BP 166, 38042 Grenoble cédex, France}

\author{A.D. Kent}
\affiliation{Department of Physics, New York University, 4 Washington Place, New York, New York 10003}

\author{G.W. Stupian and M.S. Leung}
\affiliation{Electronics and Photonics Laboratory, The Aerospace
Corporation, P.O. Box 92957, Los Angeles, CA 90009}

\begin{abstract}
The Hall response function of symmetric and asymmetric planar Hall
effect devices is investigated by scanning a magnetized tip above
a sensor surface while simultaneously recording the topography and
the Hall voltage. Hall sensor geometries are tailored using a Focused
Ion Beam, in standard symmetric and new asymmetric geometries. With
this technique we are able to reduce a single voltage probe to a narrow
constriction 20 times smaller than the other device dimensions. We
show that the response function is peaked above the constriction,
in agreement with numerical simulations. The results suggest a new
way to pattern Hall sensors for enhanced spatial resolution. 
\end{abstract}
\maketitle

\section*{Introduction}

Hall effect devices are versatile magnetometers operating in a broad
range of conditions, such as at high temperature and at high magnetic
field. They are routinely used in industry as linear and rotary motion
detectors and for power sensing. In the area of fundamental physics
Hall magnetometers have recently been used for high sensitivity and
spatial resolution magnetic imaging~\cite{chong01,oral96}, for studies
of nanoscale ferromagnetic particles~\cite{kent94} and single molecule
magnets~\cite{bokacheva00,delbarco02}. New hybrid devices combining
Hall sensors with ferromagnetic materials that behave globally as
bipolar switches are also of growing interest~\cite{johnson97,monzon97}.
The availability of 2D low electron density and high mobility materials,
hence large Hall resistance and signal to noise ratio, make the use
of Hall probes based techniques attractive especially when combined
with microfabricated devices. 

These applications require high spatial resolution, excellent field
sensitivity and good magnetic coupling between sample and sensors.
One natural way to achieve this is to reduce the dimensions of the
sensor active area defined by the intersection of the current and
voltage leads. However, as the width of these leads is reduced in
size their resistance increases. Higher lead resistance increases
device Johnson noise and power dissipation. Hence, up to now, the
design of Hall sensors is a compromise between high spatial resolution
and good signal to noise ratio. Although numerous studies have been
done to optimize the geometry of Hall devices in an homogeneous magnetic
field~\cite{popovic91}, investigation of the Hall effect in inhomogeneous
fields has been carried out only in the standard cross shaped geometry~\cite{bending97}. 

In this paper we propose a new device geometry, based on the careful
analysis of the Hall effect, that enhances the sensor spatial resolution
without \emph{increasing} lead resistances. Indeed, we show that high
spatial resolution Hall probes can be obtained by reducing the size
of one \emph{single} voltage sensing lead while keeping the other
probe dimensions macroscopic. Specifically, we consider the Hall effect
in the diffusive transport regime, in which the carrier mean free
path is smaller then the smallest sensor dimension. In contrast to
the ballistic transport regime, in which for low magnetic fields the
Hall response is proportional to the average magnetic field in the
sensitive area, in the diffusive regime the Hall response depends
strongly on local field inhomogeneities. Numerical simulations have
been carried out in our group that show that highly asymmetric Hall
probes have sensitive areas localized around the smallest sensing
voltage lead~\cite{liu98}. These results have recently been confirmed
by other groups~\cite{cornelissens01}. Here we present an experiment
that measures directly the Hall response function (HRF) of Hall sensors
with symmetric and asymmetric geometries and shows that they have
similar spatial resolution.

\section*{Experiment}

The Hall sensors were fabricated in Bi by photolithography. The \( 2500\,  \)\AA~thick
Bi polycristalline thin films were thermally evaporated on a 200 \AA~Ti
adhesion layer on sapphire substrates. Photolithography was used to
pattern the films into conventional cross like patterns. Although
numerous geometries were used, in this article we present only results
based on symmetric \( 2\mu  \)m\( \times 2\mu  \)m and asymmetric
\( 10\mu  \)m\( \times 5\mu  \)m current by voltage probes. Bi thin
films were characterized at 300 K by measuring the resistivity \( R_{\Box }\sim 10\, \Omega  \)
and Hall coefficient in a homogeneous magnetic field \( R_{H}\sim 0.5 \)~\( \Omega  \)/T.
Because of fluctuating deposition conditions and impurity rates, these
values vary a little from sample to sample. The mean free path, evaluated
using the drude model is about \( \lambda _{f}\sim 0.1\, \mu  \)m
setting the lower bound for probe dimensions in order to be in the
diffusive regime. The Hall effect was shown to be linear in homogeneous
magnetic fields until \( 2.5 \) T. The sign of the Hall coefficient
\( R_{H}=V_{H}/I \) changes between room temperature and liquid nitrogen
temperature. The strong dependence of the Hall coefficient on temperature
is due to the non-trivial transport mechanism in Bi that involve several
pockets of electrons and holes on the Fermi surface with temperature
dependant mobilities. For an isotropic material and two types of carriers,
the Hall coefficient is : \( R_{H}=(p\mu ^{2}_{p}-n\mu ^{2}_{n})/(e\times (p\mu _{p}+n\mu _{n})^{2}) \)
and can change sign~\cite{popovic91}. Thus the change of sign in
the Hall coefficient of Bi probes is not surprising and indeed has
been noticed earlier by several authors~\cite{boffoue00,butenko97,buxo80,kochowski78,inoue74}. 

All the local magnetic measurements presented in this work were made
at room temperature, and no attention was given to the actual sign
of the Hall coefficient. The key points for these studies are the
value of \( R_{H} \) at room temperature and the linearity of the
Hall effect for small magnetic fields. 

\begin{figure}
\begin{center}
\subfigure[]{\resizebox*{7cm}{!}{\includegraphics{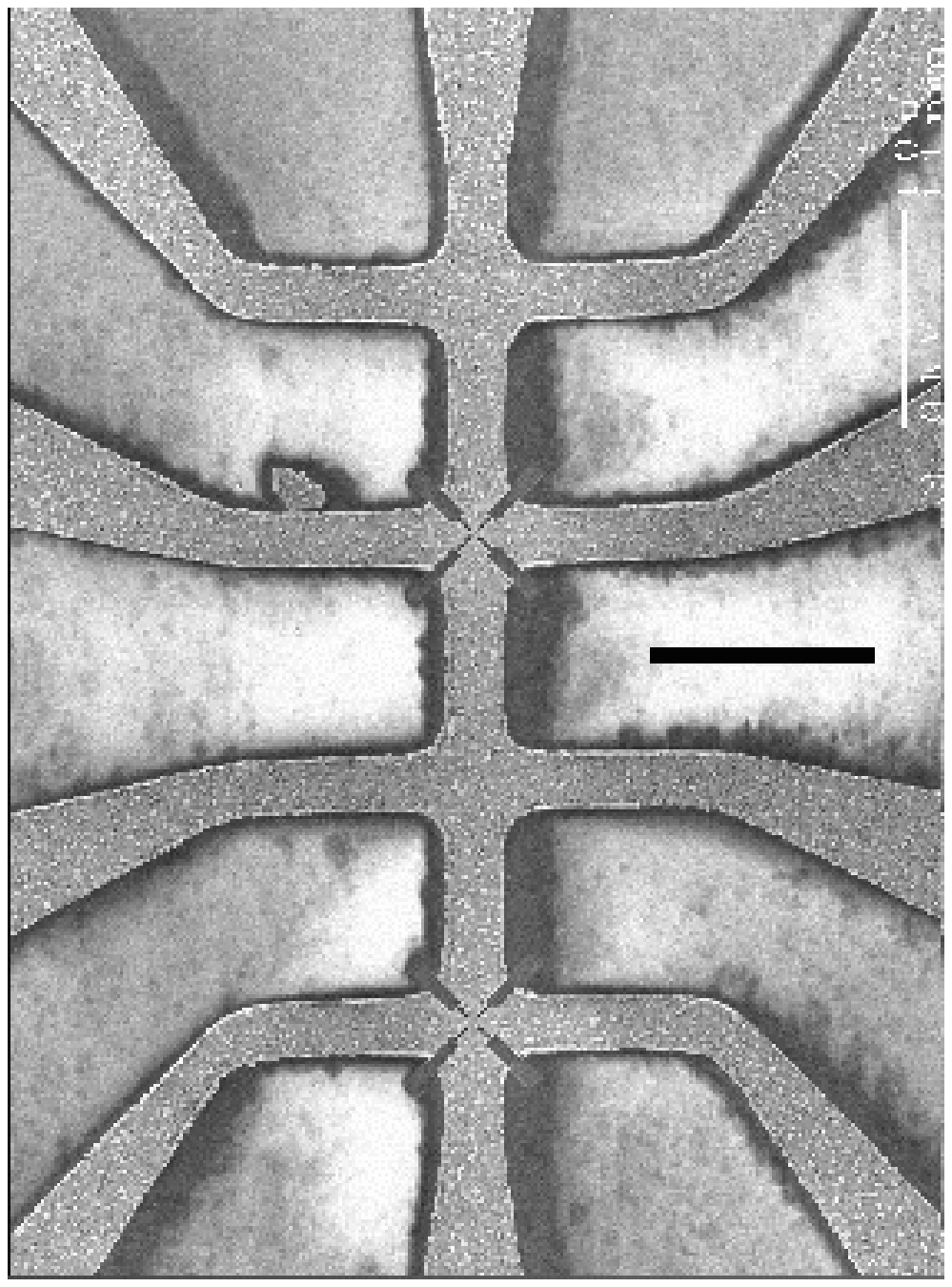}}} 
\hspace{0.25in}
\subfigure[]{\resizebox*{7cm}{!}{\includegraphics{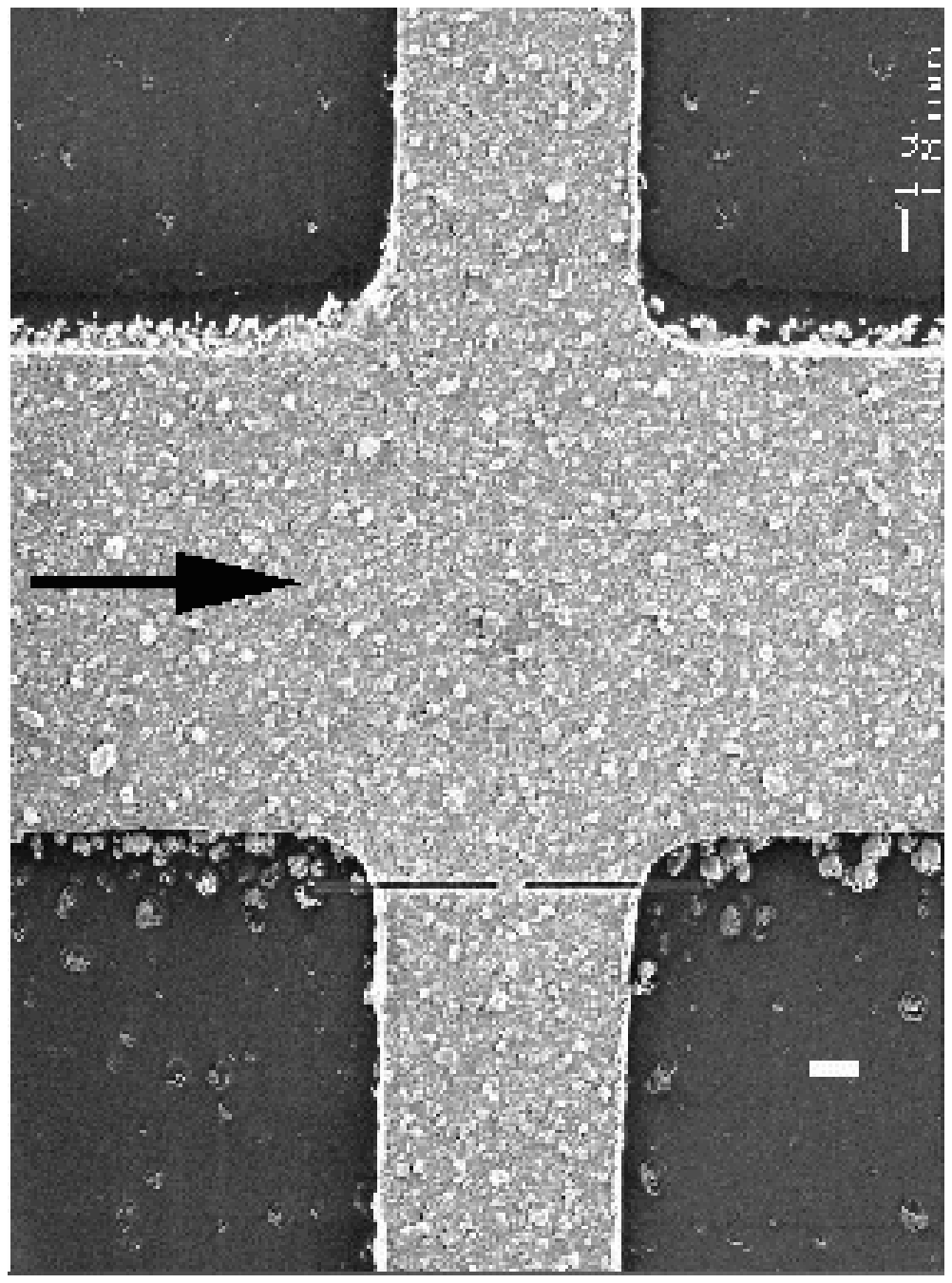}}} 
\end{center}
\caption{\label{SEM}SEM micrographs of standard cross shaped Hall probes
whose geometry has been tailored with FIB. a) nominal \protect\protect\( 2\mu \mbox {m}\times 2\mu \mbox {m}\protect \protect \)
geometry symmetrically reduced to \protect\protect\( 0.5\mu \mbox {m}\times 0.5\mu \mbox {m}\protect \protect \)
(bottom) and \protect\protect\( 0.25\mu \mbox {m}\times 0.25\mu \mbox {m}\protect \protect \)
(third cross). Black bar in figure is\protect\protect\( \, 10\mu \protect \protect \)m
in length. b) nominal \protect\protect\( 10\mu \mbox {m}\times 5\mu \mbox {m}\protect \protect \)
geometry asymmetrically reduced to \protect\protect\( 10\mu \mbox {m}\times 0.5\mu \mbox {m}\protect \protect \).
The black arrow shows the direction of the excitation current. White
bar\protect\protect\( \, =1\mu \protect \protect \)m.}
\end{figure}

We focus now on probe geometries. Numerical computations~\cite{liu98,bending97,hlasnik66,ibrahim98,cornelissens01}
show that the Hall response is highly dependent on the field distribution.
The HRF~; i.e. the Hall voltage measured as a function of the position
of a small \( \delta  \)-function like magnetic field ; was found
to be highly peaked above the voltage leads~\cite{liu98,cornelissens01}.
A design where one voltage lead is reduced to submicron sizes while
other dimensions of the probe are kept macroscopic, should thus give
a highly peaked HRF on the constricted lead and thus a high spatial
resolution sensor. To verify this result experimentally we designed
extremely asymmetric probe geometry where~: i) current and voltage
leads have different dimensions and ii) one voltage lead is reduced
to a sub-micron size. To produce such probes we used a Focused Ion
Beam to tailor the lead geometry. Indeed with such a technique it
is possible to cut the Bi films and substrate at very precise locations
with a line width smaller than 0.1 \( \mu  \)m. The FIB was used
both to design asymmetric probes by cutting only one voltage lead
and symmetric probes by cutting all four leads in a {}``clover leaf''
pattern as shown on fig.~\ref{SEM}. The leads could have been further
constricted but other problems such as robustness against electrical
shocks occur. It must also be kept in mind that the transport has
to be in a diffusive regime. 

After constructing our Hall sensor we place a localized magnetic field
with a sub-micron resolution at different positions above the surface
of the probe and record the Hall voltage as a function of the field
position. To measure the HRF we used the Hall probes as the sample
and scan a MFM tip over it while recording both the topography and
Hall voltage. The resulting image is directly the HRF convolved with
the stray field of the tip which we suppose to be very localized~\cite{streblechenko96,thiaville97}.
To remove offsets due to misalignment of voltage probes as well as
inhomogeneities in the Bi films, a bridge circuit excited by a alternating
current was used to record the Hall response. To avoid pick-up a carefully
shielded cover was added to the MFM setup and the scanning frequency
was much smaller than the excitation current to average several periods
of the excitation current for each of the 256 points of one line scan. However
some pick-up was always present and a low pass filtering of the image
yields workable data as shown on fig.~\ref{SIM} and fig.~\ref{ASIM}.

\section*{Results and discussion}

\begin{figure}
\centering 
\subfigure[]{\resizebox*{7cm}{!}{\includegraphics{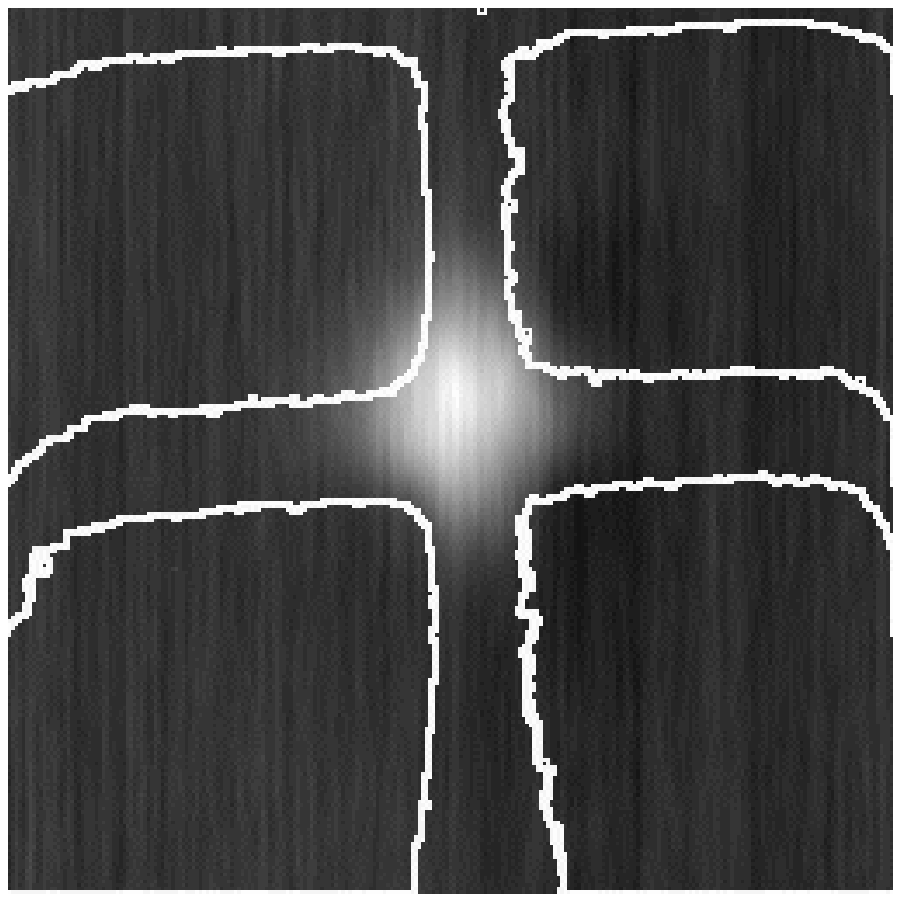}}} 
\hspace{0.25in} 
\subfigure[]{\resizebox*{7cm}{!}{\includegraphics{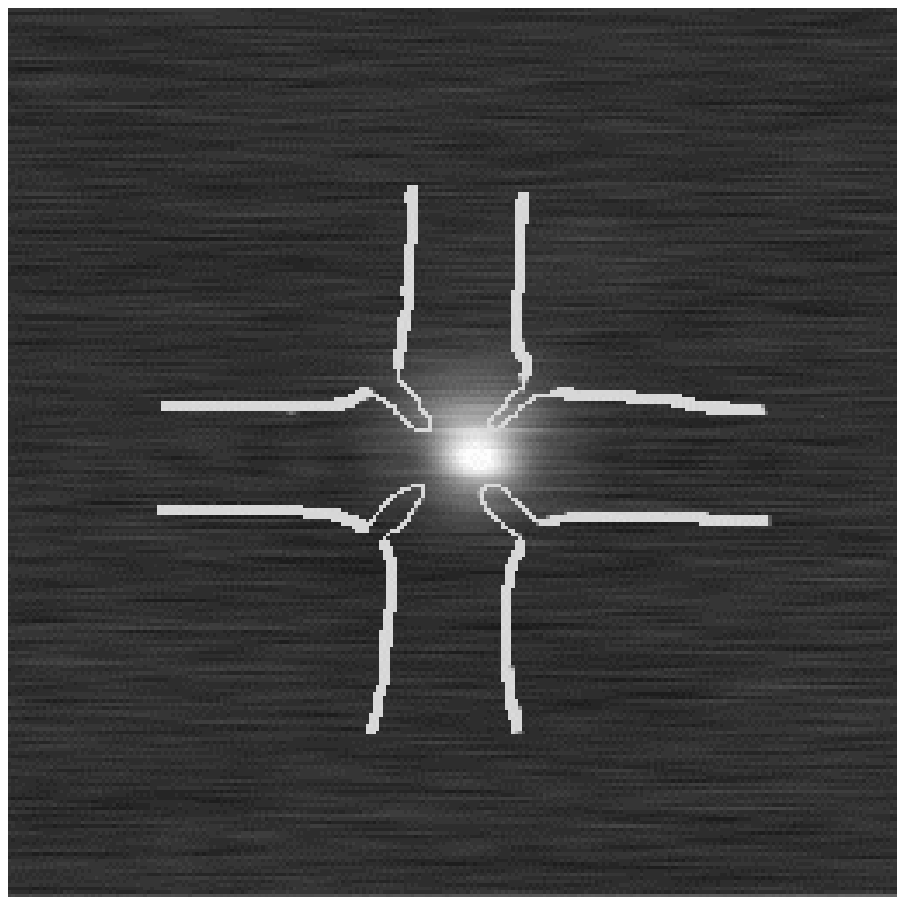}}} \par{}

\caption{\label{SIM}Hall response of symmetrical probes. The scan range is
\protect\protect\( 19\, \mu \protect \protect \)m and the greyscale
spans 2.5mT in both images. a) sensitive area is 2\protect\protect\( \mu \protect \protect \)m\protect\protect\( \times 2\mu \protect \protect \)m
; b) sensitive area is \protect\protect\( 0.5\mu \protect \protect \)m\protect\protect\( \times 0.5\mu \protect \protect \)m.
The white lines superimposed on the Hall response image outline the
edges of the topographic image. The minor shift between the topographic
and the Hall response images stems from several sources. The tilt
of the magnetized tip is one of them.}
\end{figure}

\begin{figure}
\centering 
\subfigure[]{\resizebox*{7cm}{!}{\includegraphics{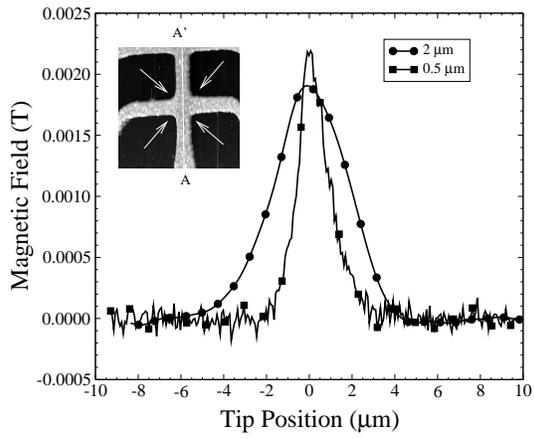}}} 
\hspace{0.25in}
\subfigure[]{\resizebox*{7cm}{!}{\includegraphics{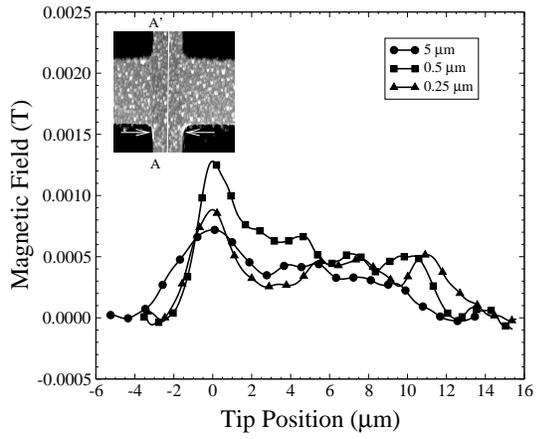}}} \par{}

\caption{\label{crossY}Cross section of the Hall response function perpendicular
to the mean current direction in the case of : a) a symmetrical probe,
and b) an asymmetrical probe. The Hall response is measured along
the white line \protect\protect\( AA'\protect \protect \). The white
arrows indicate the position of FIB cuts.}
\end{figure}

\begin{figure}
\centering 
\subfigure[]{\resizebox*{7cm}{!}{\includegraphics{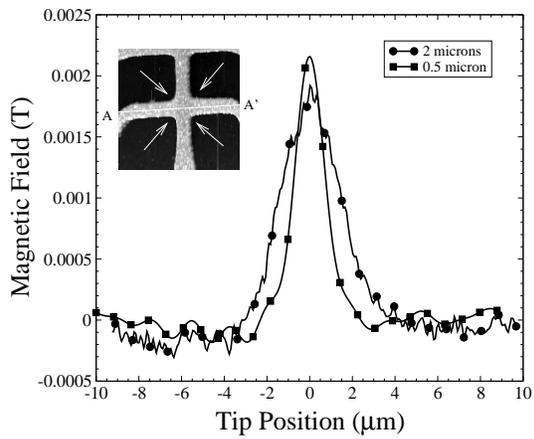}}} 
\hspace{0.25in}
\subfigure[]{\resizebox*{7cm}{!}{\includegraphics{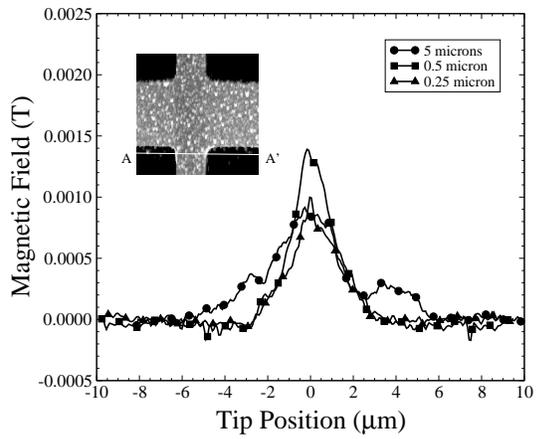}}} \par{}
\caption{\label{crossX}Cross section of the Hall response function parallel
to the average current direction. In case of a) a symmetrical probe,
and b) an asymmetrical probe. }
\end{figure}

Fig.~\ref{SIM} shows the Hall response functions of symmetric geometries
with an outline of the geometrical shape of the sensor superimposed.
The greyscales span the same magnetic field amplitude, namely \( \Delta B=2.5 \)
mT. The two geometries represented : the cross and {}``clover leaf''
are equivalent in terms of Hall effect in an homogeneous field~\cite{popovic91}.
For both experiments the measured HRF extends slightly outside the
area defined by the intersection of the current and voltage probes.
This is in agreement with numerical simulations~\cite{liu98} and
analytical studies~\cite{thiaville97} of the corresponding boundary
value problem. The difference in field amplitude between images is
due to changes in tip magnetization or tilt with respect to the sensor
plane between each image. MFM is known to have a submicron magnetic
resolution \cite{thiaville97,streblechenko96}, nevertheless the force
on the tip is an average of interactions between the tip and a magnetic
sample taking place in a finite volume of space. The suposition that
the stray field of the MFM tip is a \( \delta  \)-funtion is also
certainly not realistic. Attempts have been made to measure directly
the stray field of MFM tips with Hall bars~\cite{thiaville97}, electron
holography \cite{streblechenko96}, as well as with other techniques.
The deconvolution procedure is neither simple nor unique. However,
in a first approximation, the magnetized tip is approximated by a
magnetic dipole tilted from the \( z \)-axis. In fig.~\ref{SIM}a,
on the right side of the probe, the return field of the magnetic dipole
is observed as darker areas extending about 2 \( \mu  \)m away from
probe's edges. Such return field has never been observed and it shows
that the stray field of MFM tips extends over rather large areas.
HRF cross sections of symmetric geometries perpendicular and parallel
to the excitation current are shown on fig.~\ref{crossY}a and fig.~\ref{crossX}a
respectively. The full width at half maximum is about twice the probes
characteristic size. It is slightly larger in the perpendicular cross
section (fig.~\ref{crossY}a). The actual measured sensitive area
is thus twice as big as the surface defined by the intersection of
voltage and current probes. It is simply understood by the presence
of a nonzero current density in the voltage probes on a characteristic
length of the probes width. The current provides sufficient charge
carriers to create a local Hall potential. It is worthwhile to notice
in fig.~\ref{crossX}a the negative magnetic field corresponding
to the above mentioned dark areas. 

\begin{figure}
\centering 
\subfigure[]{\resizebox*{7cm}{!}{\includegraphics{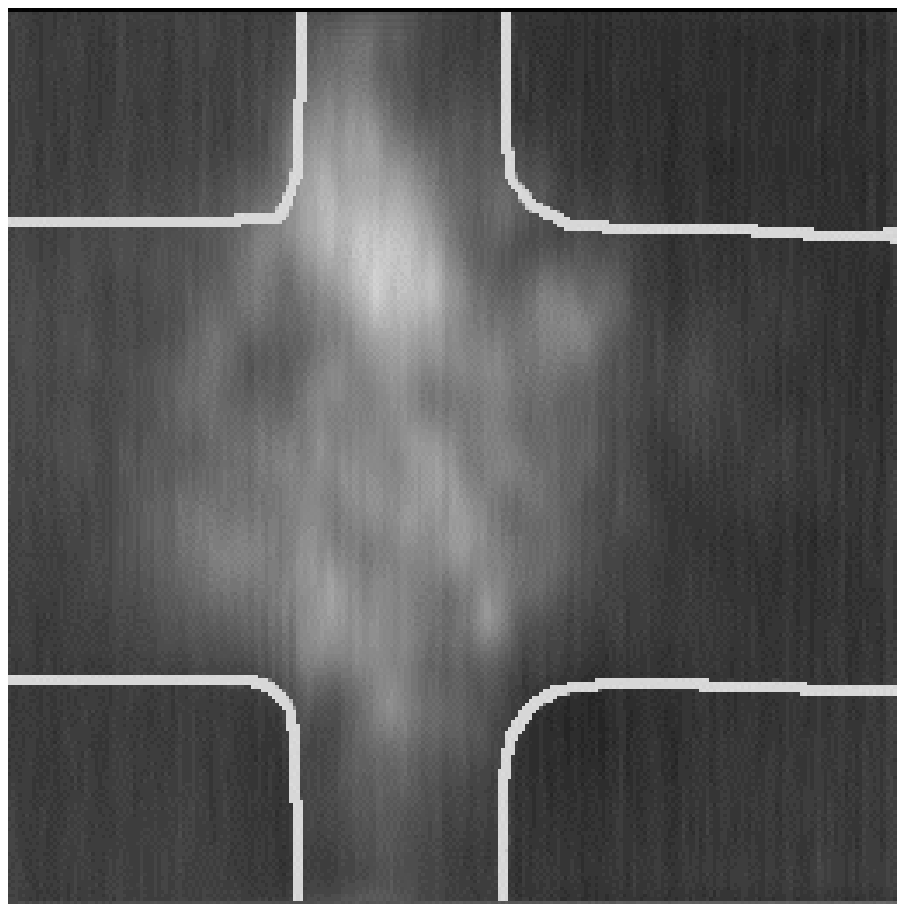}}} 
\hspace{0.25in} 
\subfigure[]{\resizebox*{7cm}{!}{\includegraphics{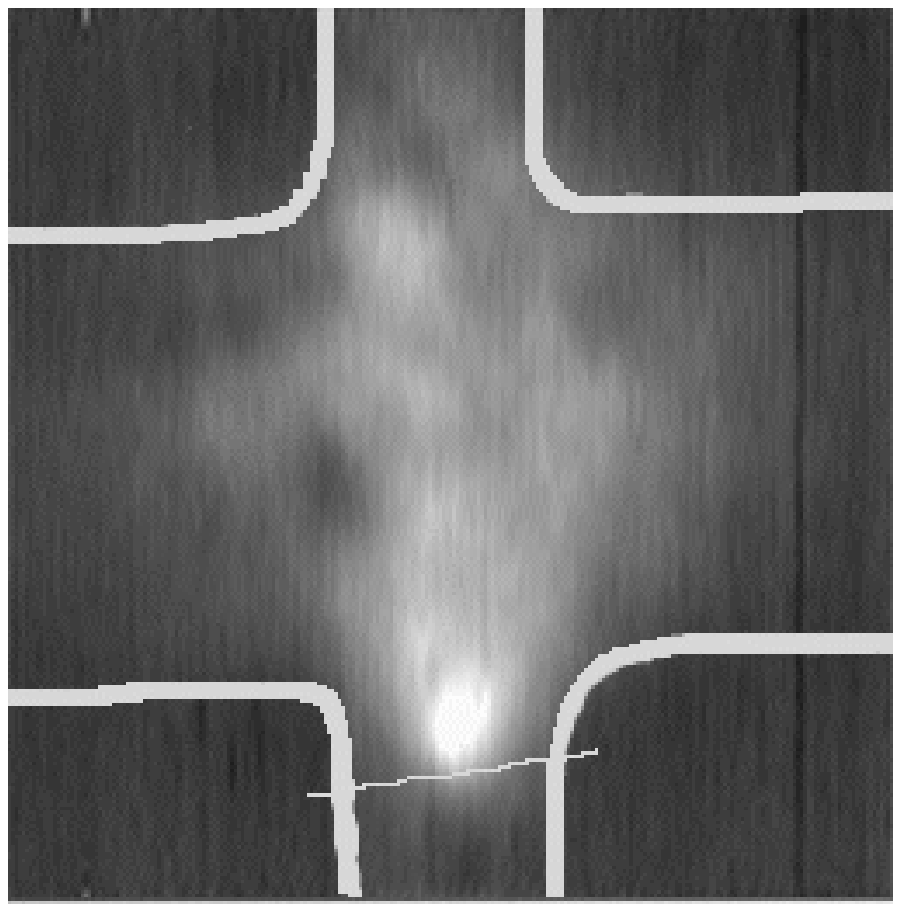}}} \par{}
\centering 
\subfigure[]{\resizebox*{7cm}{!}{\includegraphics{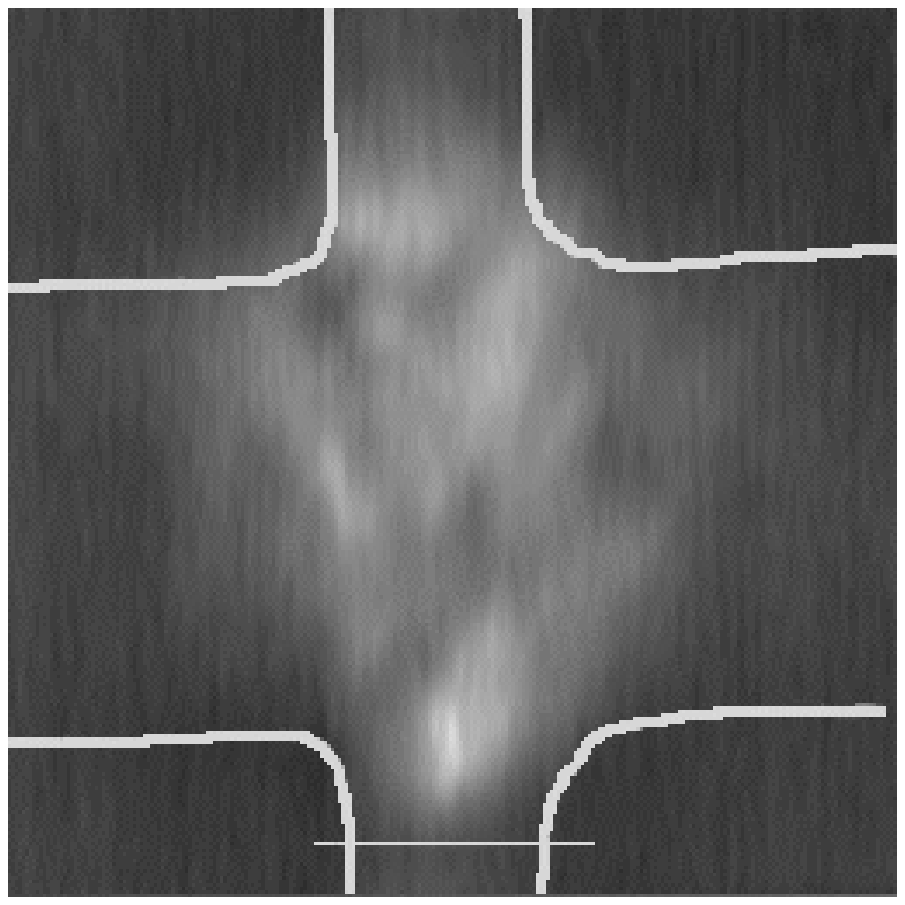}}} \par{}

\caption{\label{ASIM}Hall response of assymetrical probe geometries. The
scan range is \protect\protect\( 19\, \mu \protect \protect \)m and
the greyscales span 1.6 mT. Lithographic dimensions \protect\protect\( 10\mu \mbox {m}\times 5\mu \mbox {m}\protect \protect \)
probe a) ; asymmetrical FIB constricted \protect\protect\( 10\mu \mbox {m}\times 0.5\mu \mbox {m}\protect \protect \)
probe b) and \protect\protect\( 10\mu \mbox {m}\times 0.25\mu \mbox {m}\protect \protect \)
probe c). The thin white lines indicate the position of the FIB cut.}
\end{figure}

The HRF of asymmetric geometries are shown in fig.~\ref{ASIM}. The
greyscales have also been adjusted to span the same magnetic field
amplitude: \( \Delta B=1.6 \) mT. For the initial asymmetric \( 10\times 5\mu  \)m\( ^{2} \)
geometry (fig.~\ref{ASIM}a), although highly inhomogeneous the HRF
is peaked above the voltage probes. Inhomogeneities within the Bi
films lead to inhomogeneous current distributions and therefore inhomogeneous
HRF. The tilted dipole character of the magnetized tip also leads
to an asymmetry in Hall response between the top and bottom voltage
probes. The Hall voltage is the convolution of the magnetic field
distribution and the actual HRF, for the bottom lead the return field
of the tip is crossing the Hall sensor in the middle of the sensitive
area whereas for the top lead it is crossing the sensor in a less
sensitive area yielding a different Hall voltage. For the constricted
geometries (fig.~\ref{ASIM}b and fig.~\ref{ASIM}c) the HRF is
definitely more peaked above the constriction. The position of the
FIB cut is important since the amplitude of the HRF is different in
the two images. This is attributed again to a difference in the current
density distributions. The FIB cut made in the \( 10\mu \mbox {m}\times 0.25\mu  \)m
(fig.~\ref{ASIM}c) is further away from the current lead than the
cut made in the \( 10\mu \mbox {m}\times 0.5\mu  \)m (fig.~\ref{ASIM}b)
yielding a smaller current density at the constriction and thus a
smaller Hall response. This indicates that the Hall response is very
sensitive to the local current density. It is illustrated in fig.~\ref{crossX}b
where it is peaked above the corners of the \( 5\mu  \)m voltage
probe i.e. zones of high current densities. 

Fig.~\ref{crossY}b and fig.~\ref{crossX}b illustrate that the
reduction of one single voltage probe greatly enhances the HRF just
above the constriction and therefore the spatial resolution of the
sensor, defined as the full width half maximum of the response function.
The measured resolutions for the asymmetric probes are respectively
for the \( 5\mu  \)m, \( 0.5\mu  \)m, and the \( 0.25\mu  \)m constrictions
about \( 4\mu  \)m\( \times 10\mu  \), \( 2\mu  \)m\( \times 4\mu  \)m
and \( 2\mu  \)m\( \times 2\mu  \)m. In the case of the \( 0.5\, \mu  \)m
FIB cut, the high sensitivity area is almost the same as the one measured
for the symmetric \( 0.5\mu \mbox {m}\times 0.5\mu \mbox {m} \) cross.
The exact shape of the peak in the measured response is intimately
connected to the film inhomogeneities, the magnetic field profile,
the rounded boundaries and to the non-linearities~\cite{cornelissens01}.
It is thus difficult to model quantitatively. We therefore focus on
qualitative observations to explain our results. 

The increasing use of small scale Hall devices follow the need for
high spatial resolution imaging and good coupling between magnetometers
and magnetic materials~\cite{oral96,bokacheva00}. The constitutive
relation relating local current \( J \), electric field \( E \)
and external magnetic field \( B \) in the diffusive regime is:\begin{equation}
\label{ohm}
J=\sigma E+\mu _{H}(J\times B),
\end{equation}
 where \( \sigma  \) is the conductivity, \( \mu _{H} \) the Hall
mobility. In 2D it was shown, that the electric potential in the probe
must be a solution of the equation :\begin{equation}
\label{laplace}
\nabla ^{2}\phi =-\mu _{H}\nabla B_{z}\cdot (\nabla \phi \times \overrightarrow{z})
\end{equation}
 with the following boundary conditions : i) at conducting boundaries,
the tangential component of the electric field is zero, i.e., \( \partial \phi /\partial t=0 \)
; and ii) at insulating boundaries, the normal component of the current
is zero, i.e., \( \partial \phi /\partial n-\mu _{H}B_{z}\partial \phi /\partial t=0 \).
If \( B_{z} \) is uniform, the equation for the electric potential
becomes the Poisson equation and the solution is uniquely determined
by the boundary conditions, i.e, the fields \emph{at the boundaries}.
In the opposite case there is a source term in the equation. That
source term is the key point of our analysis. Indeed in the limit
of small magnetic fields, it acts as a distribution of electric dipoles
perpendicular to the local electric field. The dipole character of
the source term was already noticed in the 1960s~\cite{hlasnik66}.
In the case of homogeneous magnetic fields the dipole density is constant
throughout the sensor and leads to macroscopic opposite surface charge
densities at the boundaries. On the contrary, for a field distribution
smaller than the sensor characteristic dimensions, the source dipole
distribution is localized within the probe where there is a current
density and a varying magnetic field. The enhancement of the HRF in
the neighborhood of the constricted lead has two main origins. First,
the source term changes rapidly when crossing the constriction perpendicularly
to the zero field current distribution. The constriction is a good
weak link~: there is almost no current flowing through it and the
reference equipotential is in its vicinity. Second, the dipolar character
of the source term in equation~\ref{laplace} is mostly effective
nearby the sensing probe since it is short ranged in nature. These
facts, verified numerically~\cite{bending97,liu98,cornelissens01}
and now experimentally show that the geometry of planar Hall device
can be optimized to provide a high spatial resolution by scaling down
a single probe. 

In conclusion, we have measured the HRF function of planar Hall sensors
of different geometries. Specifically, we studied an asymmetric design
where one single sensing probe is reduced to a narrow constriction.
We showed that this design yields identical spatial resolution as
the standard reduced cross pattern in agreement with numerical simulation.
It validates also the dipole interpretation of the Hall effect in
inhomogeneous magnetic fields. Although FIB is not the only way to
achieve highly asymmetric geometries, it provides a versatile technique
to pattern a sensor. It should reveal its full power in fundamental
studies of magnetic nanoparticles, for example, to cut sensing probes
near the position of particles on a macroscopic sensor.

\section*{Acknowledgements}

The research at NYU was supported by the NSF Grant No. DMR-0103290
and a NSF US-France Cooperative Research Grant DMR-9729339.

\newpage 

\newpage

\newpage

\newpage

\newpage

\end{document}